**Title**: Reconfigurable swarms of colloidal particles electrophoretically drivenin nematic liquid crystals.

**Running Title**: Reconfigurable colloidal swarms in nematic liquidcrystals

**Authors**:

Sergi Hernàndez-Navarro[1,3], Pietro Tierno[2,3], Jordi Ignés-Mullol[1,3,*], and Francesc Sagués[1,3]

[1] Department of Physical Chemistry, Universitat de Barcelona.Martí i Franquès 1, 08028 Barcelona, Catalonia, Spain.

[2] Department of Structure and Constituents of Matter, Universitat de Barcelona. Avinguda Diagonal 647, 08028 Barcelona, Catalonia, Spain.

[3] Institute of Nanoscience and Nanotechnogy, Universitat de Barcelona. Catalonia, Spain.

**Abstract**:

We present experiments where anisometric colloidal microparticles dispersed in a nematic liquid crystal cell with homeotropic anchoring conditions are dynamically assembled by means of liquid-crystal-enabled electrophoresis (LCEEP) using an AC electric field perpendicular to the confining plates. A nematic host with negative dielectric anisotropy leads to a driving force parallel to the cell plates. We take advantage of the resulting gliding anchoring conditions and the degeneracy in the direction of particle motion to design reconfigurable trajectories using a photosensitive anchoring layer (azosilane self-assembled monolayer), as the particle trajectory follows the local director orientation.

**AUTHOR CONTACT INFORMATION:**

Sergi Hernàndez-Navarro:

    Email: sergihn@hotmail.com

    Phone: +34 934039252

    Fax: +34 934021231

Pietro Tierno:

    Email: ptierno@ub.edu

    Phone: +34 934034031

    Fax: +34 934021231

Jordi Ignés-Mullol (corresponding author):

    Email: jignes@ub.edu

    Phone: +34 934039290

    Fax: +34 934021231

Francesc Sagués:

    Email: f.sagues@ub.edu

    Phone: +34 934021242

    Fax: +34 934021231

**INTRODUCTION**

The assembly and transport of microscopic entities, such as particles, droplets, or microorganisms, is a permanently motivating challenge in Soft Materials science. Colloid assembly can be tuned by controlling their size[1] and shape[2], their surface chemistry[3, 4], or by suspending the inclusions in a liquid crystal[5]. Optical trapping techniques are often employed to achieve direct control over the placement of colloidal inclusions, and holographic tweezers allow to extend such control to a few hundreds of particles[6], although the technique is limited by the field of view of the optical system. Recently, new perspectives for the assembly and transport of colloids through the interplay of phoretic and osmotic mechanisms have been demonstrated [7, 8].

When electric fields are involved, Direct Current (DC) driving of colloids can lead to unwanted effects such as ion migration or electrolysis, which can be avoided by using Alternating Current (AC) fields. For AC driving to be effective, the fore-aft symmetry of individual particles must be broken. This has been realizedwith metallo-dielectric Janus particles driven in aqueous media[9], where symmetry is broken by the particle inhomogeneous surface. More recently, the use of a nematic liquid crystal (NLC) host has showed the capability to transport homogeneous solid[10-12] or liquid inclusions[13] along the local nematic director. In this Liquid Crystal Enabled Electrophoresis (LCEEP), symmetry is broken by the NLC director field around the inclusion, which must have a nonzero dipolar component in order for LCEEP propulsion to be effective. When a microscale inclusion is dispersed in a NLC, the distortion of the director field leads to the formation of defects around the particle, with a configuration that depends on the particle shape and on the anchoring conditions. For spherical particles, tangential boundary conditions result in the formation of two surface defects (boojums) leading to a quadrupolar symmetry of the director, while homeotropic boundary conditions lead either to an equatorial ring defect (quadrupolar symmetry) or to a single hyperbolic hedgehog point defect (dipolar symmetry)[5, 14, 15]. LCEEP has been demonstrated with spherical solid or liquid inclusions, and the required dipolar symmetry has been achieved by means of suitable particle surface functionalization. However, it may be challenging to prepare a sample with a large number of spherical inclusions featuring a single hedgehog point defect configuration.

In this work we use anisometric pear-shaped solid inclusions with tangential boundary conditions embedded in a NLC. Particle shape guarantees non-quadrupolar elastic distortions of the director field, thus enabling LCEEP propulsion when subjected to an external AC electric field. By employing photoelastic modulations of the NLC matrix we are able to separate particle driving and steering[16], demonstrating an unparalleled ability to control the assembly of arbitrarily large ensembles of inclusions, and to drive the formed clusters as swarms along reconfigurable paths.

**EXPERIMENTAL SECTION**

**Preparation of photosensitive glass-ITO plates**

NLC cells were prepared using 0.7mm thick microscope slides of size 15x25mm$^2$, coated with a thin layer of indium-tin oxide (ITO) with a sheet resistance of 100 Ω per square (VisionTek Systems). Photosensitive glass-ITO plates were prepared as shown in Fig. 1. In a first step, clean and dry plates where coated with a self-assembled monolayer of (3-aminopropyl)triethoxysilane (APTES, Sigma-Aldrich)[17], and subsequently rinsed with toluene, methanol (99.9%, Scharlau) and ultrapure water. Plates were then stored in a desiccator for 2h before further use. In a second step, an amide bond was formed between the amino terminal group of APTES and the custom-synthesized azobenzene 4-octyl-4'-(carboxy-3-propyloxy)azobenzene (8Az3COOH)[18]. This bond formation was typically performed[19] in a dimethylformamide (DMF, peptide synthesis grade, Scharlau) medium using pyBOP (>97%, Fluka) as the coupling agent. APTES-functionalized plates were rinsed in DMF and submerged in a Petri dish with 5ml of DMF to which 1 ml of DMF containing 0.6mg of 8Az3COOH and 1 ml of DMF containing 1.0 mg of pyBOP were added. Additionally, 6μL of N-ethyldiisopropylamine (>98%, Fluka) were added to obtain the necessary alkaline environment. The system was kept in the dark and under stirring overnight. The plates were subsequently rinsed first with DMF and then with MilliQ water, and finally dried with a stream of N$_2$. The functionalized plates were stored in the dark under Argon atmosphere until used.

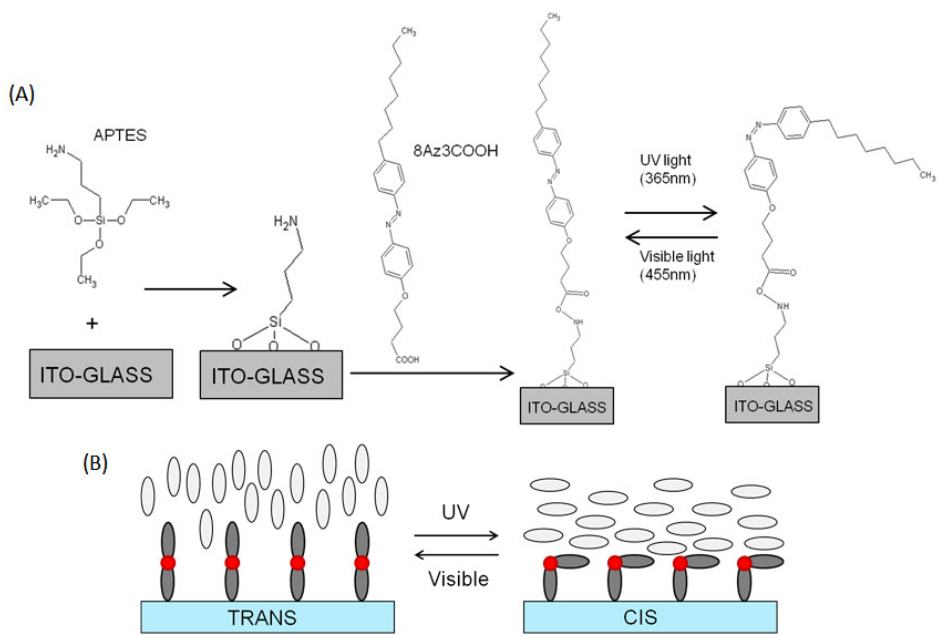

**Figure 1.** (A) Functionalization of glass substrates using APTES and 8Az3COOH, as described in the text. The result is a photosensitive azosilane anchoring monolayer. (B) *trans-cis* isomerization induced via UV-visible irradiation, and its effect on liquid crystal (LC) anchoring[20].

**Preparation of homeotropic counterplates**

Hydrophilic plates (glass-ITO plates previously coated with APTES as explained before) were spin-coated with a polyimide compound (0626 from Nissan Chemical Industries, using a 5% solution in the solvent 26also from Nissan) at 2000rpm for 10s, prebaked 1min at 80°C to evaporate the solvent and then cured for 45 min at 170 °C.

**Microscopy setup**

The experimental system was composed byan epi-illumination setup withcollimated LED sources(Thorlabs) integrated in an optical microscope (Nikon Eclipse 50iPol). Irradiation was performed using the optical path of the microscope, through its objectives. Transmitted red light (using a 645nm, 50 nm FWHM filter) was used during observations to avoid perturbation of the azosilane coating. A long-pass dichroic mirror (cutoff wavelength 405nm) was used to combine the blue (455nm, 25nm FWHM) and the UV (365nm, 10nm FWHM ) light along the same optical path. A second long-pass dichroic mirror (cutoff wavelength 505nm) prevented excitation light from reaching the camera. Tracks were imprinted on the NLC cells by moving the sample stage while the UV irradiation light is on. Estimated power densities on the samples are 1.4 W cm$^{-2}$ at 455 nm, and 0.3 W cm$^{-2}$ at 365 nm.

AC sinusoidal electric fields were applied using a function generator (ISO-TECH IFA 730) and an amplifier (TREK model PZD700), via electric contacts to LC cell plates.

Bright field and polarized optical images were captured with an AVT Marlin F-131B CMOS camera controlled with the software AVT SmartView 1.10.2. Digitized images were subsequently processed and analyzed using software packages ImageJ and IgorPro.

**LC cells preparation**

LC cells were built by gluing together an azosilane-coated plate and a homeotropic counter-plate with strips of polyethylene terephthalate (Mylar, Goodfellow) acting as spacers, obtaining cell gaps in the range 13-46µm. Polystyrene anisometric particles (pear-shaped, 2x3µm$^2$ , 3x4µm$^2$, or 8x10µm$^2$ Magsphere), initially suspended in water, were redispersed in methanol. A few microliters of the sonicated methanol dispersion were deposited on top of a small volume of the NLC (MLC-7029, Merck, room temperature properties: $\Delta\varepsilon$(1 kHz) = -3.6, $K_1$ = 16.1 pN, $K_3$ = 15.0 pN), and the solvent was let to evaporate in a desiccator. The NLC dispersion was subsequently agitated in a vortex and sonicated. LC cells werefilled by capillary action with freshly prepared dispersions.

## RESULTS AND DISCUSSION

**Photoelastic modulation of the liquid crystal cells**

Without external influences, boundary conditions of the prepared cells lead to uniform homeotropic anchoring of the NLC, with the dispersed pear-shaped particles aligning following the local director field. By irradiating with UV light, we force the azosilane monolayer to adopt the *cis* form leading to planar alignment of the local NLC director[20] (Fig. 1B, 2A, 2B).

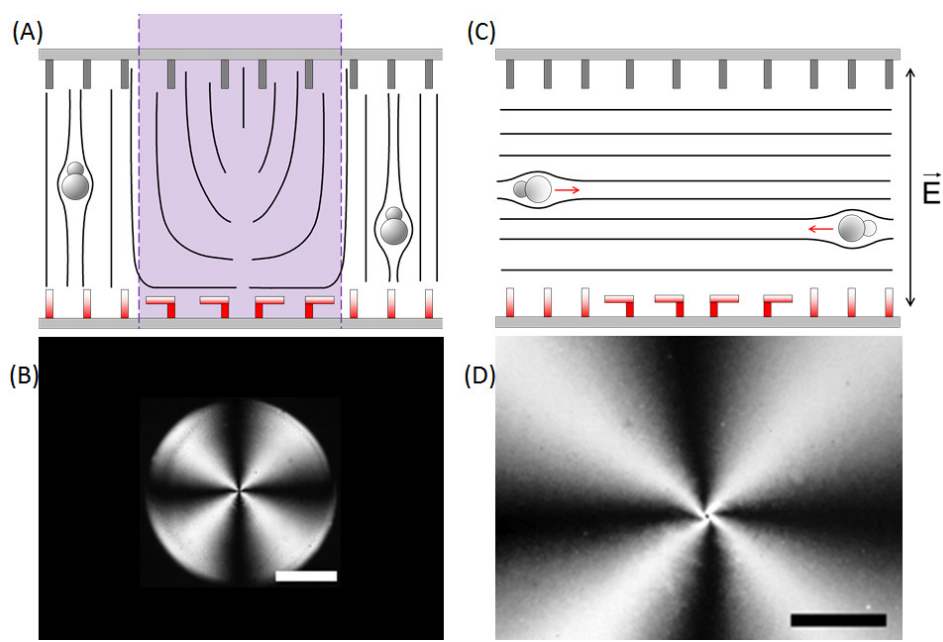

**Figure 2**. (A) Schematics of the cross section of the experimental cell upon irradiation with a UV light spot. The bottom plate is coated with the azosilane film, which is isomerized into the *cis* form, leading to hybrid anchoring conditions across the cell gap. (B) Top-view micrograph between crossed polarizers. The scale bar is 500 μm long. (C) Cross section schematics of the cell upon application of a sinusoidal electric field between the plates. Red arrows indicate the dominant direction of motion of the anisometric particles. (D) Top-view micrograph between crossed polarizers of the configuration in (C). The scale bar is 200 μm long.

The unpolarized beam of UV light has a spatial intensity profile with approximately a Gaussian shape, which leads to a radial splay arrangement of the NLC director in contact with the irradiated spot (Fig. 2A, 2B), organized around a central s=+1 point defect. Outside of the spot, the director remains perpendicular to the cell. Application of an external AC field aligns the negative dielectric anisotropy NLC parallel to the plates everywhere[21]. Because of the homeotropic anchoring, the field-induced planar alignment is energetically degenerate. However, the in-plane radial boundary conditions imposed by the irradiated spot break this degeneracy, propagating the pure splay texture radially outwards for several millimeters. The induced configuration is stable for days under an AC field, well past the half-life for thermal relaxation of the azosilane film, which is about 30 minutes in this system.

**Liquid-crystal enabled electrophoretic driving**

Upon application of the AC field, particles tumble instantaneously following the NLC director so that their long axis lays, on average, parallel to the cell plates. Simultaneously, LCEEP sets the particle into motion along the local director at a constant speed, which is a balance between electrophoretic propulsion and viscous drag. Particle speed has a leading quadratic dependence on the electric field amplitude (Fig. 3A) and features a maximum around 10 Hz for 3×4 $\mu m^2$ particles (Fig. 3B). These results agree qualitatively with the behavior of spherical solid[10] or liquid[13] inclusions driven by LCEEP under different configurations. Speed changes drastically with particle size. It is undetectably slow for the smallest particles (2×3 $\mu m^2$), and has significantly lower values that those reported in Fig. 3 for the largest particles (8×10 $\mu m^2$).

During propulsion, we find that most particles are oriented so that their symmetry axis is roughly parallel to the direction of motion, with the larger lobule ahead (Fig. 2C), although there is a significant dispersion in their orientation, as shown in Figure 3C with data using the largest particles in order to better resolve their orientation. LCEEP arises from the anisotropic ionic mobility around the inclusions, and this, in turn, is modified by the defect pattern induced by the inclusion in the surrounding NLC director. Given the particle geometry in the present case, one would expect the formation of two boojums (surface defects) along the symmetry axis. The fact that particles are dispersed in the NLC without annealing it into the isotropic phase might generate pinning on inhomogeneities present on the particles surface, leading to metastable defect arrangements. This might justify the anomalous particle orientation we observe during LCEEP motion (Fig. 3C), and the disordered packing obtained upon particle assembly (see below).

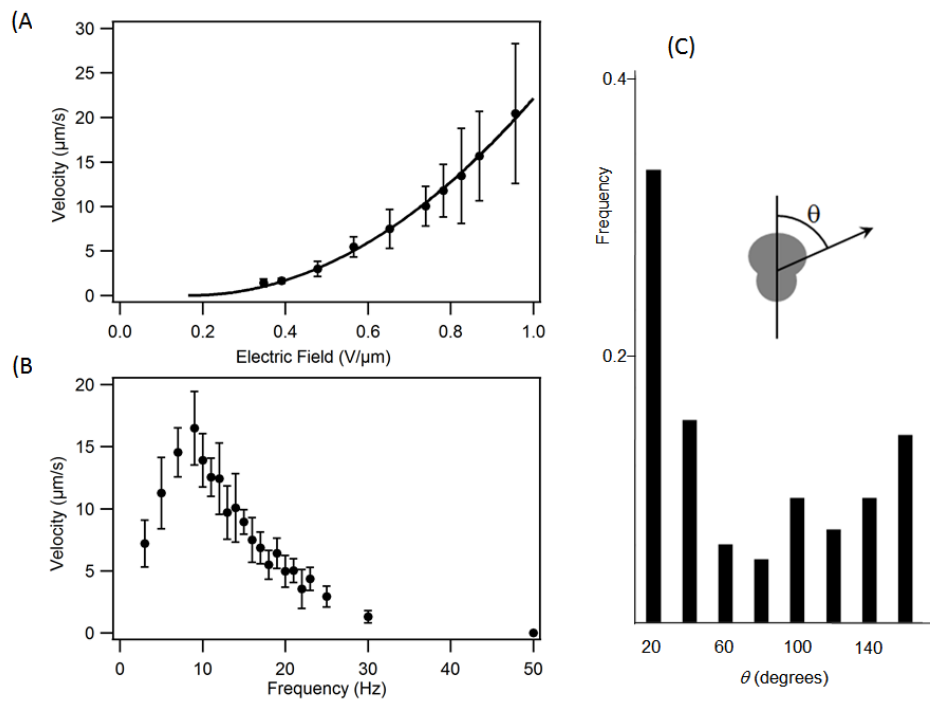

Figure 3. Experimental measurements of the dependence of the velocity of pear-shaped particles (size 3×4μm$^2$) driven by LCEEP on the electric field amplitude (A, frequency $f$= 10Hz) and frequency (B, field strength $E$ = 0.73 V/μm). The solid line in (A) is a quadratic fit. In panel C, histogram of the orientation of particles (size 8×10μm$^2$) with respect to their moving direction.

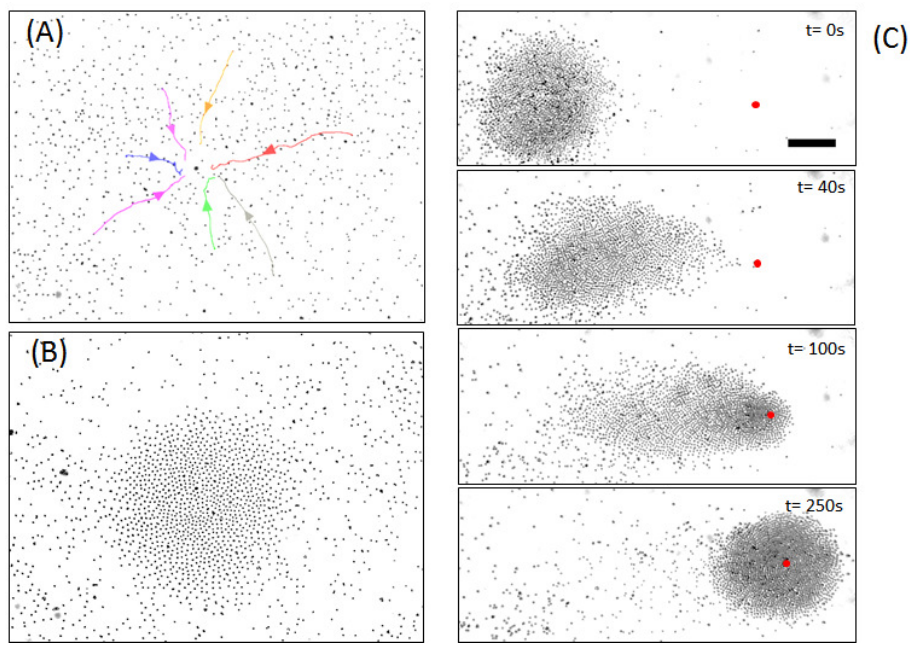

Figure 4. A radial pattern imprinted on the NLC cell by a UV light spot (Figure 2) attracts dispersed particles (A) that assemble into an arrested aster-like cluster following centripetal trajectories (B). In (C), the UV light spot has been rewritten 600 µm to the right. Upon application of the AC field (t = 0s), the cluster of particles moves as a swarm towards the center of the new spot.

**Cluster assembly and swarm transport**

The region with radial alignmentacts asa basin of attractionfor dispersed particles, which assemble around the original UV-irradiated spot into an arrested aster configuration (Fig. 4). If multiple UV spots are distributed over the sample surface, they compete as attractors, resulting in the formation of a lattice of assembled particle clusters. In the absence of an applied field, assembled clusters are stable against spontaneous disaggregation due to diffusion for weeks. This can be understood by noticing that, for the used NLC with a dynamic viscosity of about $\eta$ = 0.1 Pa s, the self-diffusion coefficient will be of order $D = 10^{-3} \mu m^2 s^{-1}$. An assembled particle cluster can be repositioned in the cell by erasing the original UV spot and scribing a new one. Upon application of the AC field, particles move as a swarm towards the new attractor, forming a new cluster around it (Fig. 4C).

As discussed above, the circular region with radial alignment induced by UV irradiation features a central s=+1 point defect. This topological charge must be balanced with a negative defect of the same magnitude to make this structure compatible with the uniform farfield NLC alignment. One would typically expect the formation of an s=-1/2 ring disclination surrounding the circular region in Fig. 2B. Upon application of the AC field, the region with radial alignment grows, and the ring disclination often collapses into an s =-1 companion point defect. The presence of such a defect is easily observed as it scatters particles during LCEEP motion, a consequence of the hyperbolic field lines that are traced by the moving particles. The placement of the companion defect is typically unpredictable, as it may remain close to the boundary of the original circular spot or it can be expelled far away. However, regular arrangements of s=+1 defects can lead to the formation of a complementary lattice of s=-1 defects, and even to secondary s=+1 defects surrounded by negative counterparts (Fig. 5). Such a patterning of aligning monolayers can lead to interesting defect healing scenarios in NLC cells with hybrid anchoring conditions[22, 23].

**Controlling the dynamic state of colloidal clusters**

The LCEEP-induced assembly of colloidal inclusions can be modified from the formation of static aster-like aggregates to rotating mills, where assembled particles rotate at a constant linear velocity around the center of the aggregate (Fig. 6). This can be achieved by taking advantage of the higher cost of splay with respect to bend distortions in this NLC, since the elastic constant $K_1$ is 7% higher than $K_3$ (see above). We begin by *erasing* the central region of a UV-irradiated spot using a smaller spot of blue light (Fig. 6). With this, in the absence of electric field, the NLC director features a homeotropic configuration both outside and inside a corona with planar radial alignment.

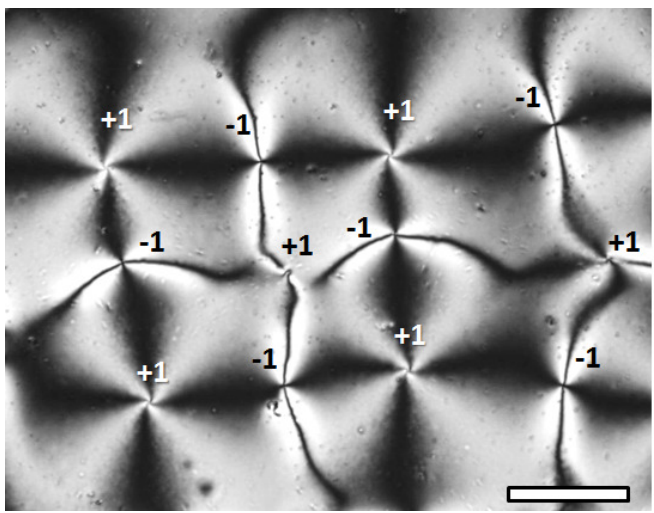

Figure 5. Micrograph between crossed polarizers of a NLC cell where a lattice of circular spots has been imprinted with a center-to-center distance of about 650 µm. Upon application of the AC field, complementary s=-1 and secondary s=+1 defects (black labels) are formed. The scale bar is 300 µm.

Upon application of the AC field, the NLC director both inside and outside the ring adopts a degenerate planar alignment because of the homeotropic anchoring conditions. As explained above for the full circular spot case (see Fig. 2), the degeneracy is broken by the radial boundary conditions both on the outer and on the inner ring boundary. The energy cost of the large splay distortions in the region inside the ring prompts the NLC director to adopt a spiral bend-splay configuration. This does not happen in the case where only a spot of UV light is applied, since the alignment on the *cis*-azosilane surface imposes a pure splay radial distortion. Nevertheless, even in this case one can observe that the region surrounding the central s=+1 defect features a bend-splay, rather than a pure bend, texture (Fig. 2D).

Using a similar protocol, the dynamic state of an already assembled cluster of particles can be reversibly and selectively switched from the arrested aster to the rotating mill configuration, thus offering an additional degree of control on this colloidal assembling process.

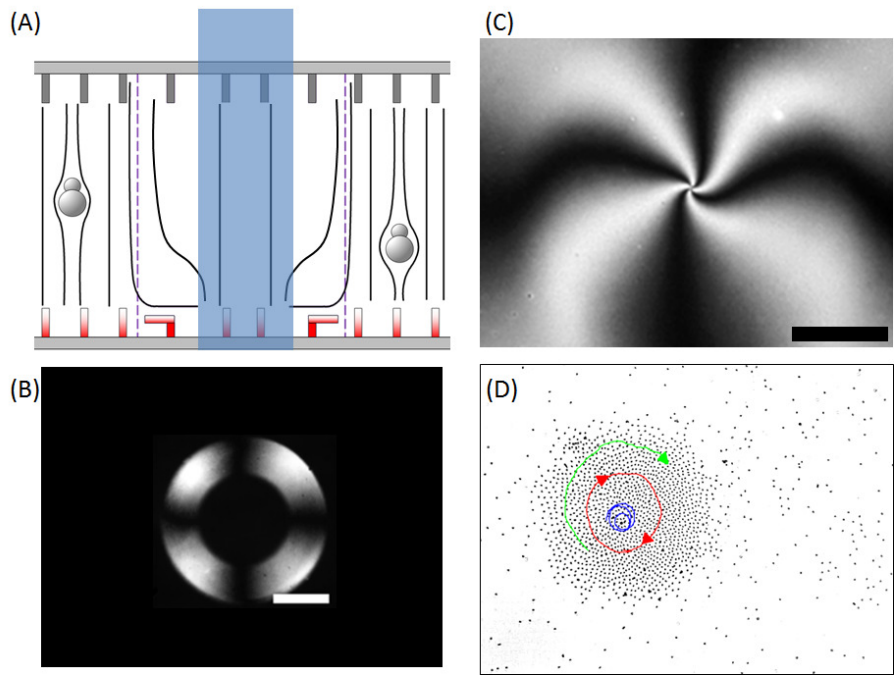

Figure 6. (A) Schematics of the cross section of a NLC cell previously irradiated with a UV spot that is now illuminated at the center with a narrower spot of blue light, which leads to the isomerization of the azosilane layer into the *trans* form. (B) Top view micrograph between crossed polarizers revealing a corona with planar radial alignment. The scale bar is 500 µm. (C) Top view of the spiral texture formed by the NLC director upon application of an AC field. The scale bar is 200 µm. (D) Particles are attracted to the center of the irradiated spot following spiral trajectories and assemble into a rotating mill. The trajectories of three different particles are tracked for the same amount of time.

**CONCLUSIONS**

In this study, we have reported a novel protocol to control the assembly of colloidal inclusions dispersed in a nematic liquid crystal layer by means of liquid-crystal enabled electrophoresis. Taking advantage of the degenerate planar anchoring obtained with a mesogen with negative dielectric anisotropy, we are able to independently control particle driving, achieved by means of an AC electric field applied across the cell gap, and particle steering, realized by means of a custom-synthesized azosilane monolayer that enables the photoelastic modulation of the NLC director field. We have shown that particles can be assembled into single or multiple clusters of arbitrary size by suitable patterns of UV light irradiation, with the possibility to reversibly address the dynamical state of individual clusters. Clusters of particles can be driven as swarms with this technique by scribing arbitrary paths with light on the photosensitive alignment layer. Inclusions can be of different nature, either solid or liquid, since the only requirement for LCEEP driving is on the topology of the distortion caused by the inclusions on the NLC matrix. The reconfigurability and remote addressability displayed by the reported protocol paves the way for novel strategies in lab-on-a-chip applications. One could even envisage more complex geometries in the NLC matrix that include disclination lines. Since LCEEP drives the inclusion along the local director field, linear defects might be used to generalize the colloidal transport along two-dimensional paths reported here into transport strategies that involve localization in the third dimension. From a more fundamental perspective, the reported assembly process poses challenging questions on the effective interaction among colloidal particles embedded in the NLC matrix. For instance, assembled clusters display a large steady-state interparticle distance (see Fig. 4B and 6D), which indicates the existence of long range repulsive forces that can actually be experimentally tuned by means of the frequency of the applied AC field.

**Acknowledgements**

We thank Patrick Oswald for the polyimide compound, and Joan A. Farrera for the azobenzene synthesis and assistance in chemical functionalization. We acknowledge financial support by MICINN (Project numbers FIS2010-21924C02, FIS2011-15948-E) and by DURSI (Project no. 2009 SGR 1055). S.H.-N. acknowledges the support from the FPU Fellowship (AP2009-0974). P.T. further acknowledges support from the ERC starting grant "DynaMO" (No. 335040) and from the "Ramon y Cajal" program (No. RYC-2011-07605).